\documentclass[12pt,aps,eqsecnum,nofootinbib,floatfix]{revtex4}

\usepackage{graphics}
\usepackage{graphicx}
\usepackage{epstopdf}
\usepackage{subfigure}
\usepackage{amssymb,amsmath}
\usepackage{CJK}
\usepackage{indentfirst}
\usepackage{amsmath}

\def\be{\begin{equation}}
\def\ee{\end{equation}}
\def\bea{\begin{eqnarray}}
\def\eea{\end{eqnarray}}
\def\pt{\partial}
\def\nn{\nonumber}

\begin{document}

\title{$Q-\Phi$ criticality and microstructure of charged AdS \\ black holes in $f(R)$ gravity}

\author{Gao-Ming Deng$^{1,2}$\footnote{e-mail:~gmd2014@emails.bjut.edu.cn}, Yong-Chang Huang$^{1,3}$}

\medskip

\affiliation{\footnotesize $^1$ Institute of theoretical physics, Beijing University of Technology, Beijing 100124, China\\
                           $^2$ School of physics and space science, China West Normal University, Nanchong 637002, China\\
                           $^3$ CCAST(World Lab.), P.O.Box 8730, Beijing 100080, China}

\begin{abstract}

The phase transition and critical behaviours of charged AdS black holes in $f(R)$ gravity with a conformally invariant Maxwell (CIM) source and constant curvature are further investigated. As a highlight, this research is carried out by employing new state parameters $(T, Q, \Phi)$ and contributes to deeper understanding the thermodynamics and phase structure of black holes. Our analyses manifest that the charged $f(R)$-CIM AdS black hole undergoes a first order small-large black hole phase transition, and the critical behaviours qualitatively behave like a Van der Waals liquid-vapor system. However, differing from the case in Einstein's gravity, phase structures of the black holes in $f(R)$ theory exhibit an interesting dependence on gravity modification parameters. Moreover, we adopt the thermodynamic geometry to probe the black hole microscopic properties. The results show that, on the one hand, both the Ruppeiner curvature and heat capacity diverge exactly at the critical point, on the other hand, the $f(R)$-CIM AdS black hole possesses the property as ideal Fermi gases. Of special interest, we discover a microscopic similarity between the black holes and a Van der Waals liquid-vapor system.
\bigskip


PACS Number(s):{~04.50.Kd,~04.70.-s,~04.70.Dy,~64.60.-i}

\end{abstract}

\maketitle

\tableofcontents
\bigskip

\section{Introduction} \label{section1}

There is now a general consensus that our Universe is experiencing an accelerating expansion. To give a persuasive theoretical interpretation of such puzzling acceleration, $f(R)$ theory \cite{Sotiriou:2008rp,DeFelice:2010aj} has been advanced as a promising candidate. In contrast to the standard theory of General Relativity (GR), $f(R)$ theory, where the Lagrangian density is a generic function of Ricci scalar $R$, undoubtedly serves as a class of more general theories. GR can be recovered as the special case $f(R)$=$R$. What's more, $f(R)$ theory is proved to be conformally equivalent to GR plus a scalar field \cite{Whitt:1984pd,Barrow:1988xh,Maeda:1988ab,Wands:1993uu}. For such theory model, by performing a conformal transformation mapping from the Jordan frame to Einstein frame \cite{Faraoni:2000wk}, the analysis of the background cosmological evolution can be significantly simplified. One can read the recent review \cite{Capozziello:2011et} to gain a comprehensive understanding. As a popular modified theory, it has been widely applied in gravitation and cosmology \cite{Capozziello:2007ec,Nojiri:2006gh,Nojiri:2010wj,Motohashi:2012wc} so far. Especially, $f(R)$ theory has succeeded in mimicking the historical evolution of the cosmology. Since it is considered as an attractive modification of GR, some distinctive properties from Einstein's gravity are highly expected. Further studies of $f(R)$ theory, for example, black hole solutions and thermodynamic properties \cite{delaCruzDombriz:2009et,Moon:2011sz,Cembranos:2011sr}, are of particular interest. It is worth mentioning that, in contrast to Bekenstein-Hawking radiation, Addazi et al. \cite{Addazi:2016prb,Addazi:2017cim} have remarkably explored $f(R)$ black holes' anti-evaporation phenomena, where the radius of black hole horizon could increase. Interestingly, taking into account this intriguing effect, some primordial black holes may survive in current universe. However, in the presence of a matter field in $f(R)$ gravity, the field equations get complicated and it's intractable to construct an exact black hole solution. Sheykhi creatively assumed the curvature $R$=$R_0$=$const$ and considered the traceless energy-momentum tensor for the matter field, eventually achieved a higher dimensional black hole solution \cite{Sheykhi:2012zz} in $f(R)$ gravity with a conformally invariant Maxwell (CIM) source. Motivated by these facts, this present work will take the four dimensional case as an example to further investigate the properties of the charged $f(R)$-CIM AdS black holes \cite{Sheykhi:2012zz}, such as phase transitions, critical behaviours and microstructure. Effects of the gravity modification parameters are expected to be presented.

Black hole thermodynamics has been a long-standing and fascinating subject \cite{Bardeen:1973gs,Bekenstein:1973ur,Wald:1999vt,Li:2005nm,Wu:2011im,Deng:2014bta,Liu:2015tqa},
which provides interesting clues to the underlying structure of quantum gravity. In particular, promoted by the AdS/CFT correspondence \cite{Maldacena:1997re,Witten:1998zw}, the investigation of black hole phase transitions in AdS spacetimes has received popular attention \cite{Cvetic:1999ne,Sahay:2010wi,Belhaj:2012bg,Gunasekaran:2012dq,Poshteh:2013pba,Dehghani:2014caa,Zhang:2014uoa,Chaturvedi:2015kpa,Mahapatra:2016dae,Bhattacharya:2017hfj}. The pioneering study of the transitions can be traced back to the well-known Hawking-Page phase transition \cite{Hawking:1982dh}, which describes a first-order phase transition between Schwarzschild AdS black holes and the thermal AdS space. Inspired by the AdS/CFT, Witten has explained such an appealing phenomenon as a confinement/deconfinement transition in the dual quark gluon plasma \cite{Witten:1998zw}. For Reissner-Nordstr\"{o}m (RN) AdS black holes, Chamblin et al. \cite{Chamblin:1999tk} first realized that the phase transition amazingly resembles a Van der Waals liquid-vapor system. Later, Banerjee et al. \cite{Banerjee:2011au} checked the Ehrenfest scheme of arbitrary dimensional RN AdS black holes in detail and verified that the transition occurring at critical point is the second order Van der Waals-like phase transition. Meanwhile, Dolan et al. also discovered similar behaviours in the rotating case \cite{Caldarelli:1999xj,Dolan:2011xt}. In Ref.\cite{Kubiznak:2012wp}, Kubiznak and Mann further illustrated the analogy of $4d$ charged AdS black hole with the Van der Waals fluid and affirmed the coincidence of their critical exponents. Up to now, it is found that the Van der Waals-like phase transition prevails such as in BTZ black holes \cite{Hendi:2017mgb}, Gauss-Bonnet black holes \cite{Cai:2013qga,Zeng:2016aly}, Born-Infeld black holes \cite{Zou:2013owa}.

It is noteworthy that a great many previous analyses of black hole phase structure in AdS spacetimes were based on a popular proposal \cite{Kastor:2009wy}, which identifies the cosmology constant $\Lambda$ and its conjugate quantity as thermodynamic variables \cite{Cvetic:2010jb,Dolan:2012jh,Caceres:2015vsa}, i.e., pressure $P$ and volume $V$, respectively. As a result, temperature $T$ and the newly-introduced variables $P$, $V$ were chosen as state parameters to delineate black hole phase structures and critical behaviours in various AdS spacetime backgrounds \cite{Hendi:2012um,Suresh:2014pra,Johnson:2014pwa,Xu:2014kwa,Ma:2013aqa,Wei:2015iwa,Guo:2015waa,Fernando:2016sps,Pradhan:2016dun}. Nevertheless, aside from the variables $P$ and $V$, whether there exits another pair of parameters available to describe black hole phase transitions, such as electric charge $Q$ and its  potential $\Phi$ for charged black holes. If so, some underlying thermodynamic properties are expected. The answer is positive. For charged black holes, electric charge $Q$ is an intrinsic and physical character. In fact, Chamblin and Wu \cite{Chamblin:1999tk,Wu:2000id} have successfully put this idea into practice and intuitively displayed the Van der Waals-like behaviours of RN black holes in $Q-\Phi$ plane. A well-defined statistical description still unclear, analyzing from diverse perspectives has an important physical significance. After all, the more directions we view a problem from, the more comprehensive and transparent picture we capture. Consequently, in this paper, we attempt to choose $(T, Q, \Phi)$ as state parameters to describe the physical picture of phase transitions for the charged $f(R)$-CIM AdS black hole \cite{Sheykhi:2012zz}. As a highlight, our approach provides one more alternative perspective other than $(T,P,V)$ description for one to better understand the rich critical phenomena of charged AdS black holes.

Furthermore, we would like to detect the microstructure of the charged $f(R)$-CIM AdS black holes \cite{Sheykhi:2012zz} using thermodynamic geometry. The thermodynamic geometry was initiated by Weinhold \cite{Weinhold:1975mg} and has been developed by Ruppeiner et al. \cite{Ruppeiner:1983zz,Ruppeiner:1995zz,Aman:2005xk}. Ruppeiner argued that Riemannian geometry could give insight into the underlying statistical mechanical system \cite{Ruppeiner:1995zz}. In this picture, the divergence of scalar curvature $\mathbb{R}$ is closely related to phase transitions. By calculating $\mathbb{R}$, microscopic interactions and the resulting macroscopic phase transition can be linked. Nowadays, it is widely believed \cite{Janyszek:1990rg,Oshima:1999rs,Banerjee:2010da,Mo:2013sxa} that $\mathbb{R}$ serves as an elegant measure of global fluctuations in thermodynamic systems caused by interactions. Recently, Ruppeiner further pointed out \cite{Ruppeiner:2010tc} that the sign of the curvature may be an indicator of the interactions among the microstructure of thermodynamic system. For instance, the curvature $\mathbb{R}<0$ signals the repulsive interaction \cite{Janyszek:1990rg,Oshima:1999rs}, which corresponds to ideal Fermi gases, while $\mathbb{R}>0$ signals the attractive interaction, which corresponds to ideal Bose gases.

This paper proceeds as follows. In Sec.\ref{section2}, we briefly review the charged $f(R)$-CIM AdS black hole \cite{Sheykhi:2012zz} and its thermodynamics. In Sec.\ref{section3}, the local stability and phase transitions are analyzed. We concentrate on exploring the heat capacity and exhibiting phase structures of the black holes. Sec.\ref{section4} is devoted to employing $(T, Q, \Phi)$ as new state parameters to investigate the critical behaviours in detail.
In Sec.\ref{section5}, we adopt thermodynamic state space geometry to further study the phase transitions and underlying properties of microstructure. Sec.\ref{section6} ends up with some discussions and conclusions.

\section{Charged $f(R)$-CIM AdS black holes and thermodynamics} \label{section2}

To facilitate further investigating the charged $f(R)$ AdS black hole \cite{Sheykhi:2012zz} sourced by a conformally invariant Maxwell (CIM) field, we first briefly review its some basic thermodynamic properties. $f(R)$ theory containing a CIM source contribution can be defined via the action
\be
\mathcal{S} = \int_{\mathcal{M}}d^{n}x\sqrt{-g}\left[ R+f(R)-(F_{\mu \nu}F^{\mu \nu })^p \ \right], \label{eq1}
\ee
where $f(R)$ is an arbitrary function of scalar curvature $R$, $F_{\mu \nu }=\pt_{\mu }A_{\nu}-\pt_{\nu }A_{\mu }$ the electromagnetic field tensor, $p$ a positive integer. In general, it's a challenge to solve the complicated field equations. Recently, Hendi \cite{Hendi:2009sw} has discussed the relation between $f(R)$ gravity and Einstein-CIM source, and obtained black hole solutions for the special case $f(R)=0$. In fact, for the Lagrangian (\ref{eq1}), one can straightforward check that the trace of energy-momentum tensor ${T^{\mu}}_{\mu}$ vanishes when spacetime dimension $n=4p$ and enable the field equations to be solved. Assuming the curvature $R$=$R_0$=$const$ and considering the static spherically symmetric metric ansatz, Sheykhi successfully constructed a charged $f(R)$-CIM AdS black hole solution \cite{Sheykhi:2012zz}. In this paper, we focus our attention on the case of four dimensions which can be described by
\be
ds^2 = -N(r)dt^2 + \frac{dr^2}{N(r)} + r^2 \left( d\theta^2 + sin^2\theta d\phi^2 \right), \label{eq2}
\ee
with
\be
N(r) = 1-\frac{2m}{r}+\frac{q^2}{\xi r^2}-\frac{R_0}{12}r^2\,, \quad  \xi=1+f'(R_0)\,, \label{eq22}
\ee
where the constant curvature $R_0 = 4 \Lambda$ takes negative values in asymptotically AdS spacetime, the ``prime'' stands for the derivative with respect to $R$. It is worth pointing out that parameter $\xi\neq 0$, and this solution would reduce to standard $4d$ RN AdS black hole \cite{Wu:2000id} if specifying $\xi=1$ (i.e., $f'(R_0)$ vanishes) and replacing $R_0$ with $4\Lambda$ in metric function $N(r)$. What's more, parameters $m$ and $q$ are related to the black hole mass $M$ and charge $Q$ respectively as follows
\be
M = \xi m \,, \quad   Q = \frac{q}{\sqrt{\xi}}\,. \label{eq3}
\ee
The electric potential, entropy and Hawking temperature can be calculated as
\be
\Phi = \frac{\xi Q}{r_+} \, ,\quad \quad S =\pi \xi r_+^2\,,   \label{eq4}
\ee
\be
 T = \frac{N'(r_+)}{4 \pi} = \frac{- R_0 r_+^4 + 4 r_+^2 - 4 Q^2 }{16 \pi  r_+^3}\,, \label{eq5}
\ee
where $r_+$ is the outer horizon radius (the largest root of $N(r)$=$0$) of the $f(R)$-CIM AdS black hole. We express $\Phi$ and $T$ as the function of $Q$ for later use. It can be verified that these quantities obey the first law of black hole thermodynamics \cite{Sheykhi:2012zz}
\be
dM = TdS+\Phi d{Q}\,.  \label{eq6}
\ee

\section{Local stability and phase transition} \label{section3}

\begin{figure}[htbp]
\centerline{ \scalebox{0.90}{\includegraphics{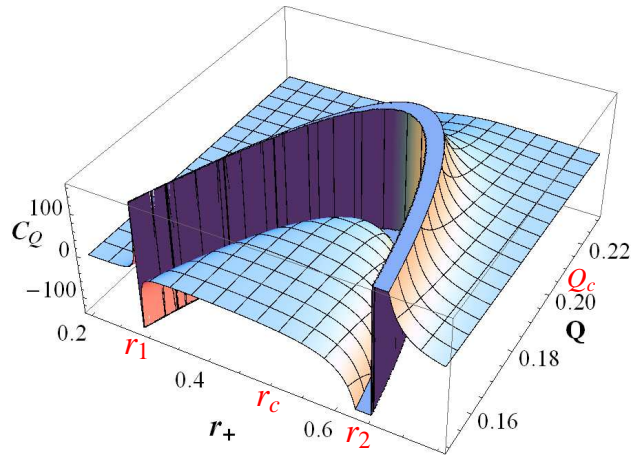}} }
\caption{(color online)\,. Heat capacity $C_Q$ varying with $r_+$ and $Q$ for fixed $\xi$=$1.5$, $R_0$=$-8$.}
\label{CRQ3Ddiag}
\end{figure}

\begin{figure*}[htbp]
  \begin{center}
    \subfigure[~$C_Q$ vs. $r_+$ for $Q<Q_c$]{\label{CqR1diag}\includegraphics[width=2.5in,height=2.0in]{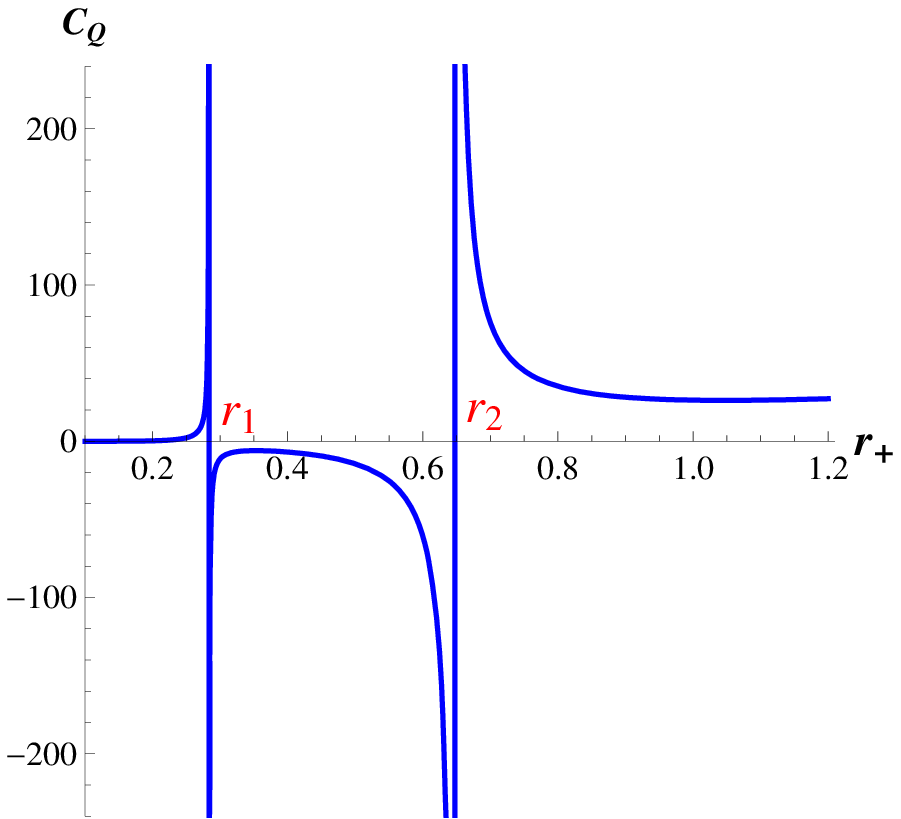}} \quad
    \subfigure[~$C_Q$ vs. $r_+$ for $Q=Q_c$]{\label{CqR2diag}\includegraphics[width=2.5in,height=2.0in]{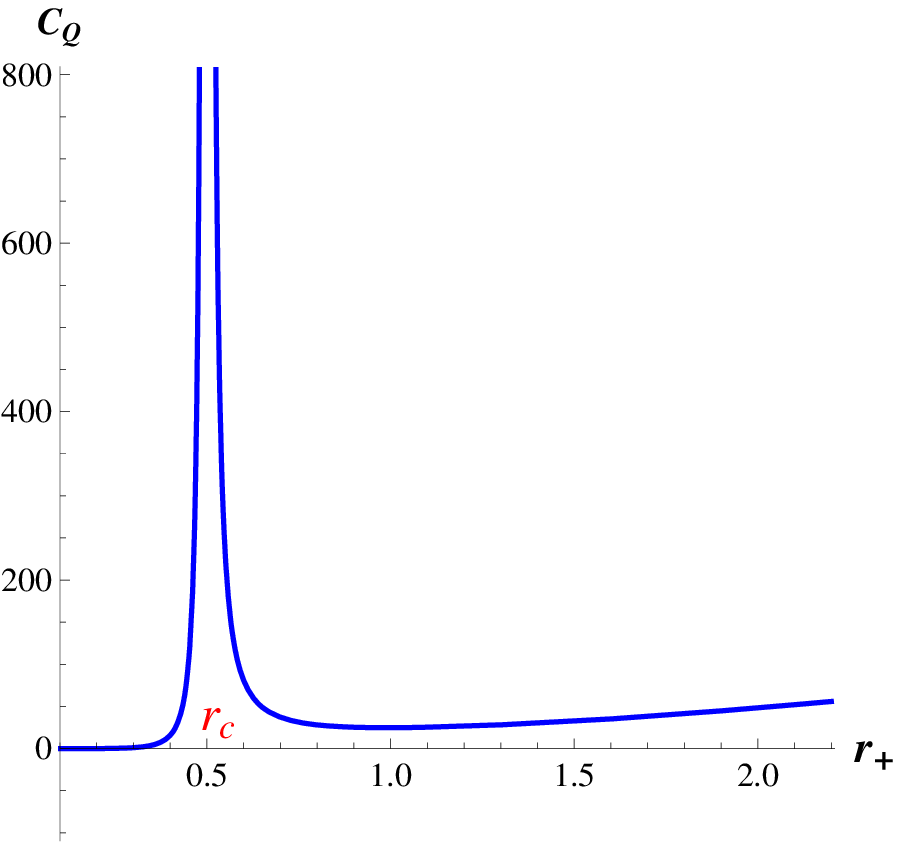}} \\
    \subfigure[~$C_Q$ vs. $r_+$ for $Q>Q_c$]{\label{CqR3diag}\includegraphics[width=2.5in,height=2.0in]{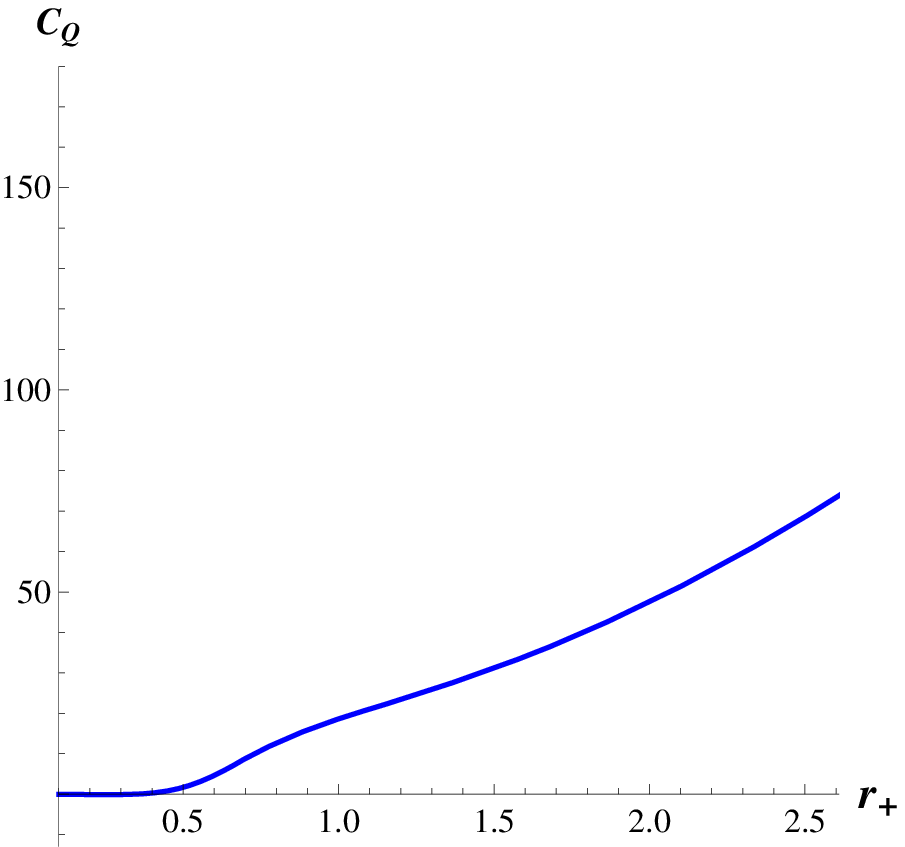}} \quad
    \subfigure[~$Q$ vs. $r_+$]{\label{QRdiag}\includegraphics[width=2.2in,height=2.0in]{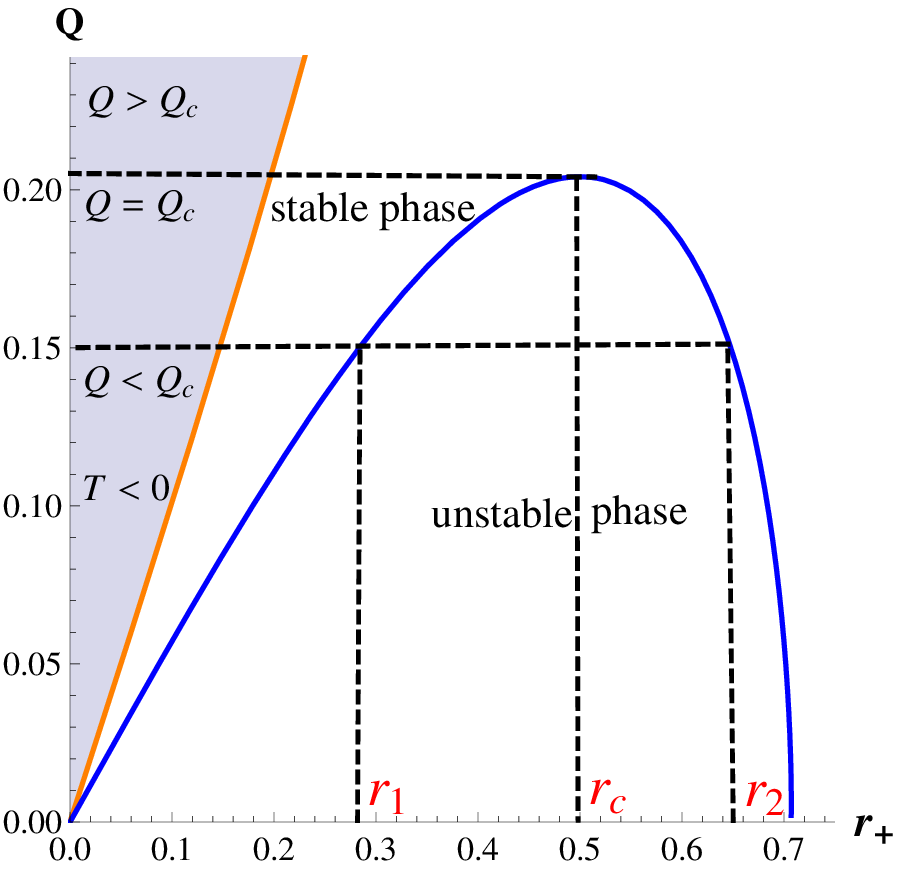}}
    \caption{(color online)\,. The divergent behaviour of the heat capacity as a function of $Q$ and $r_+$. Here we set $\xi$=$1.5\,,R_0$=$-8$. The negative temperature (shadow) region in (d) corresponds to an unphysical black hole case, and we rule it out from our consideration.}
  \label{CQRProjectiondiag}
  \end{center}
\end{figure*}

To begin with, we explore the heat capacity $C_Q$ at constant charge. With Eqs. (\ref{eq4}) and (\ref{eq5}) in hand, the heat capacity $C_Q$ can be calculated as
\be
C_Q = T \left(\frac{\pt S}{\pt T} \right)_{Q} = \frac{2\xi \pi  r_+^2 \left(R_0 r_+^4 - 4 r_+^2 + 4 Q^2\right)}{R_0 r_+^4 + 4 r_+^2-12 Q^2}\,. \label{eq7}
\ee
In general, the local stability is measured by the heat capacity. In detail, the positive heat capacity ensures a black hole stably exist, while the negative one signals that the black hole will disappear when suffering a small perturbation. It is noticeable that the heat capacity $C_Q$ diverges when the denominator
\be
R_0 r_+^4 + 4 r_+^2-12 Q^2 = 0\,. \label{eq8}
\ee
Solutions for $r_+$ degenerate into single one when
\be
1 + 3 Q^2 R_0 = 0.  \nn
\ee
The corresponding charge and horizon radius are determined by
\be
Q_c = \frac{1}{\sqrt{-3 R_0}}\,, \quad r_c = \sqrt{\frac{2}{-R_0}}\,. \label{eq9}
\ee
Behaviours of the heat capacity can be witnessed intuitively in Fig.\ref{CRQ3Ddiag}.
Seen from this picture, $C_Q$ undergoes a striking divergence at the points $r_1$ and $r_2$. Specially, for a small value of the charge $Q$, with the increase of horizon radius $r_+$, the heat capacity $C_Q$ first soars to positive infinity at $r_+=r_1$, whereafter dramatically increases from negative infinity to a finite negative value and falls back to negative infinity again at $r_+=r_2$. Ultimately, it decreases from positive infinity to a finite positive value. Closer observations can be acquired in Fig.\ref{CQRProjectiondiag}.
As is depicted by Fig.\ref{CqR1diag}, apparently, $(C_Q,\,r_+)$ plane is partitioned into three regions. $C_Q$ is positive for $r_+<r_1$ and $r_+>r_2$, while it is negative for $r_1<r_+<r_2$. Therefore both the small radius region $r_+<r_1$ and the large radius region $r_+>r_2$ are thermodynamically stable while the intermediate radius region $r_1<r_+<r_2$ is unstable. Consequently, the phase transition takes place between small black hole and large black hole. $r_1$ and $r_2$ correspond to the phase transition points of the charged $f(R)$-CIM AdS black hole. It is at the point $r_+=r_1$ that a small stable black hole with $C_Q>0$ changes to an intermediate unstable one with $C_Q<0$, and at the point $r_+=r_2$ that an intermediate unstable black hole evolves into a large stable one. As the charge $Q$ increases, two divergent points approach to each other and merge into one at $r_+=r_c$ (see Fig.\ref{CqR2diag}). The charge coupled with $r_c$ corresponds to the local maximum $Q_c$ (shown in Fig.\ref{QRdiag}), namely, critical phase transition point. But for $Q>Q_c$, the fact that the heat capacity always remains positive (see Fig.\ref{CqR3diag}) indicates that the black hole is locally stable and there will be no phase transition occurs.

\section{$Q-\Phi$ criticality and Van der Waals-like behaviours} \label{section4}

Research on the local stability reveals that the heat capacity $C_Q$ is plagued by divergences and there suffers from phase transitions between small and large black holes. In this section, we proceed to further study the critical behaviours. Interestingly, we attempt to employ $(T, Q, \Phi)$ as the state parameters.

Combining Eqs.(\ref{eq4}) and (\ref{eq5}), we express $Q$ as a function of $\Phi$ and $T$, that is
\be
Q = \frac{2 \Phi}{R_0 \xi^2}\left[-4\pi \xi T + \sqrt{16 \pi^2 \xi^2 T^2 - R_0(\Phi^2 - \xi^2)} \right],  \label{eq10}
\ee
where $R_0$ is the constant curvature which takes negative values and $\xi$ the gravity modification parameter characterizing $f(R)$ theory. Resorting to the state equation (\ref{eq10}) and derivatives
\be
\left(\frac{\pt Q}{\pt \Phi}\right)_T = 0 \, ,\quad \quad \left(\frac{\pt^2 Q}{\pt \Phi^2}\right)_T = 0\,,    \label{eq11}
\ee
one can establish the critical point as
\be
T_c = \frac{\sqrt{-R_0}}{3\sqrt{2}\pi} \, ,\quad Q_c = \frac{1}{\sqrt{-3 R_0}}\,, \quad \Phi_c = \frac{\xi}{\sqrt{6}}\,. \label{eq12}
\ee
Obviously, the same critical quantities as those in Eq.(\ref{eq9}) are presented. It implies that discussion from the $Q$-$\Phi$ perspective is consistent with the heat capacity analysis. It is particularly interesting that, for charged $f(R)$-CIM AdS black holes, critical point (\ref{eq12}) exhibits a strong dependence on the parameters $R_0$ and $\xi$.

\begin{figure*}[htbp]
  \begin{center}
    \mbox{
      \subfigure[~$\xi=1.5,\, R_0=-8$]{\label{QP1a1diag}\includegraphics[width=2.6in,height=2.0in]{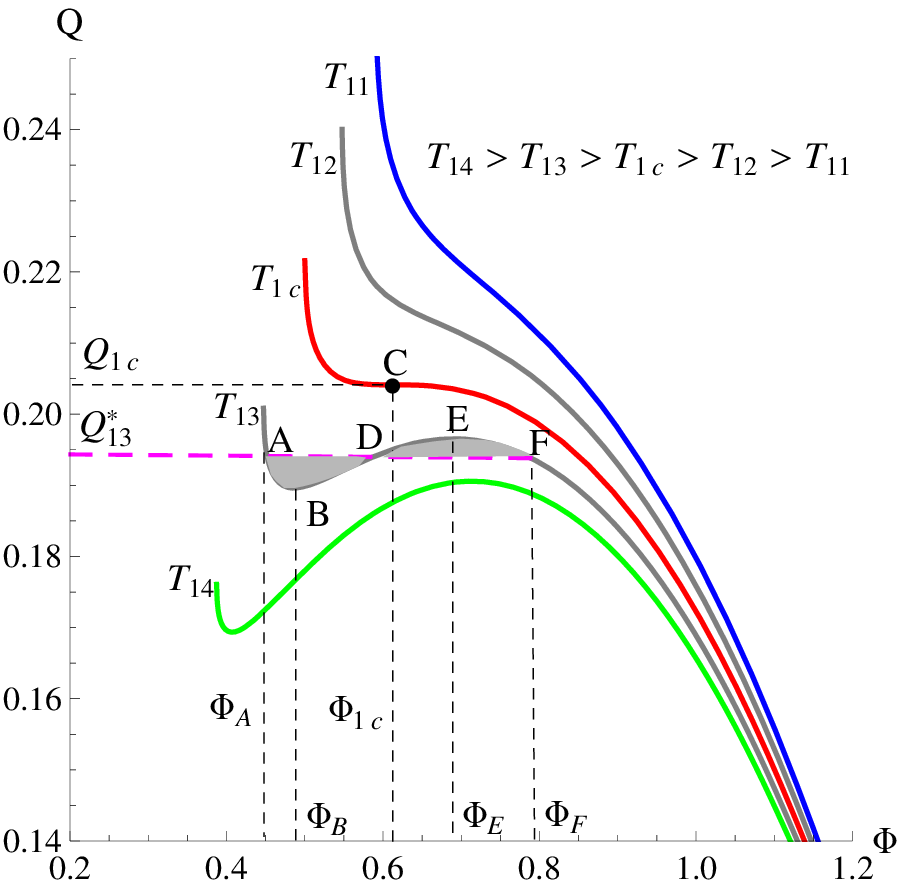}} \quad
      \subfigure[~$\xi=1.5,\, R_0=-10$]{\label{QP1b1diag}\includegraphics[width=2.6in,height=2.0in]{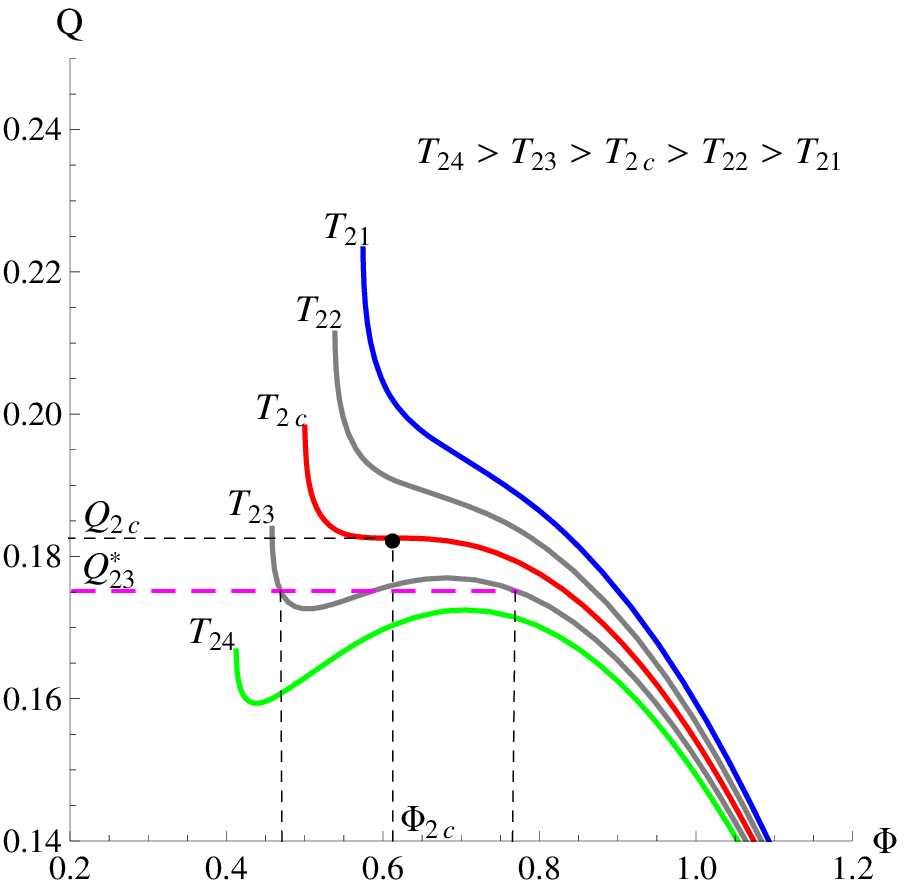}}
         }
    \mbox{
      \subfigure[~$\xi=0.2,\, R_0=-8$]{\label{QP1a2diag}\includegraphics[width=2.6in,height=2.0in]{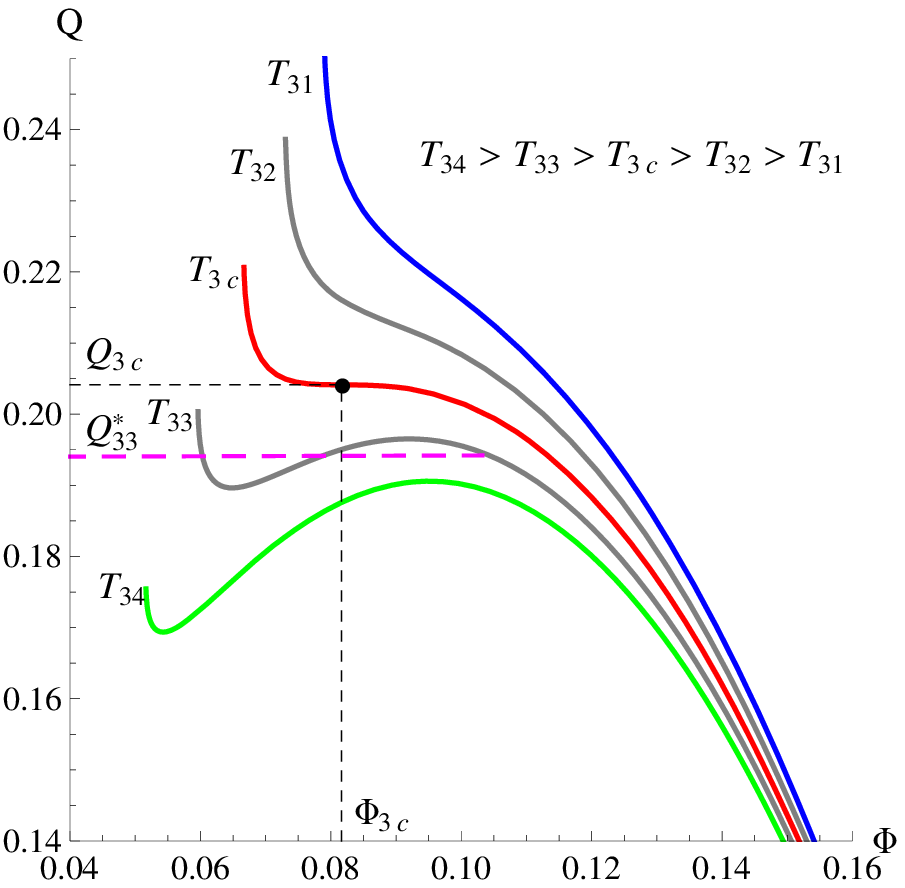}} \quad
      \subfigure[~$\xi=1,\, R_0=-10$]{\label{QP1b2diag}\includegraphics[width=2.6in,height=2.0in]{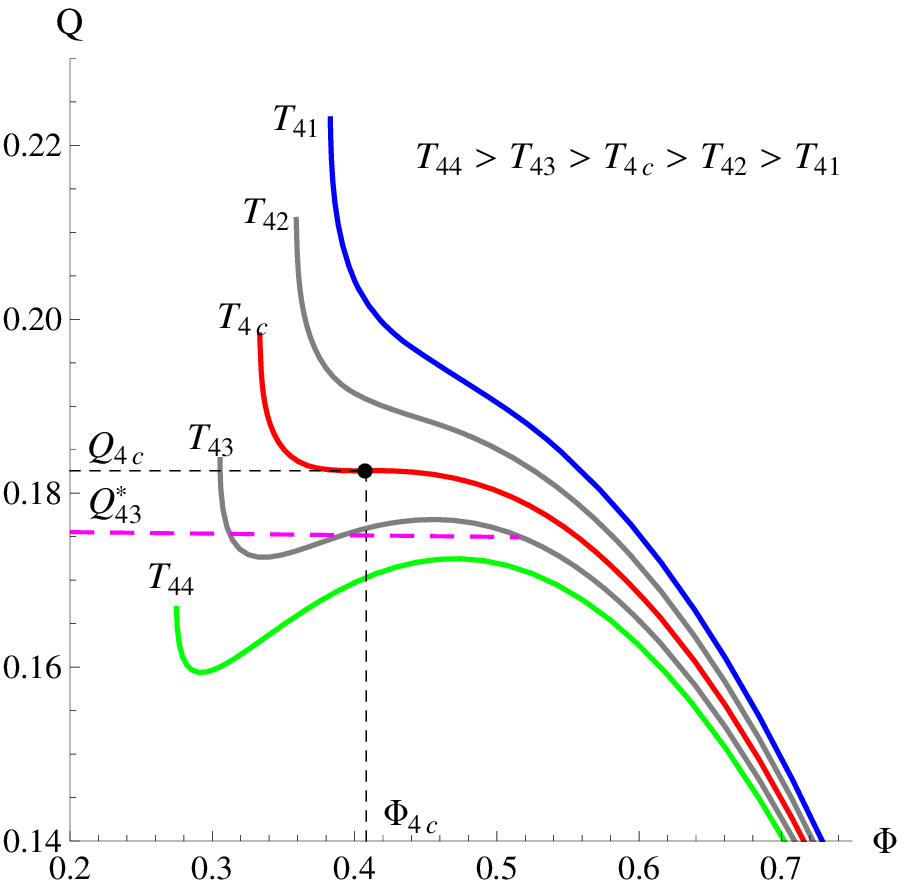}}
         }
    \caption{(color online)\,. $Q$-$\Phi$ diagram (isotherm) for different $\xi$ and $R_0$. Critical entities are labelled with subscript $``c"$. Temperature of isotherms in each picture decreases from bottom to top. The critical isotherm is depicted by the red curve and tangent to isocharge line at the critical point. The limit case $\xi=1$ in (d) coincides with that of standard $4d$ RN AdS black hole \cite{Ma:2016aat}.}
  \label{QPhidiag}
  \end{center}
\end{figure*}

To better understand the phase transition of the $f(R)$-CIM AdS black hole, one may appreciate the qualitative behaviour of isotherms in the $Q$-$\Phi$ plane in Fig.\ref{QPhidiag}.
It is quite evident that the $Q$-$\Phi$ diagram of the charged $f(R)$-CIM AdS black hole strikingly resembles the $P-V$ diagram of a Van der Waals liquid-vapor system. When temperature is higher than critical value, the isotherms are characterized by the intriguing Van der Waals oscillation. Each isotherm has a local minimum and a local maximum (e.g., the points $B$ and $E$ for the isotherm $T=T_{13}$ in Fig.\ref{QP1a1diag}). For the segments $\Phi<\Phi_B$ and $\Phi>\Phi_E$, the slope $(\frac{\pt Q}{\pt \Phi})_T < 0$ and the black hole stays in a thermally stable phase. While, within the range $\Phi_B<\Phi<\Phi_E$, the result that the slope remains positive implies that the black hole can not exist stably. In addition, within a certain range of charge $Q$, a charge corresponds to three possible black hole solutions for the same temperature $T$. It is unphysical. Fortunately, according to Maxwell's equal area law, one can adjust the oscillation curve segment $ABDEF$ by replacing it with the isocharge line (purple dashed) segment $AF$ which the (shadow) area below and above are equal. As is illustrated in Fig.\ref{CQRProjectiondiag}, for $Q<Q_c$, there exists an unstable phase with $C_Q<0$ interpolating between the locally stable small black hole and large black hole. Therefore, phase transition between a small black hole and a large black hole takes place along the isotherm. It is worth pointing out that, since a larger $r_+$ corresponds to a smaller $\Phi$ for a fixed $Q$, for the isotherm $T=T_{13}$, the curve segments for $\Phi<\Phi_A$ and $\Phi>\Phi_F$ correspond to locally stable large and small black hole phases respectively, and the region $\Phi_A<\Phi<\Phi_F$ denotes a small-large black hole coexistence phase, which is thermally unstable.

With temperature decreasing to the critical value, the oscillating segment squeezes to an inflection point $C$ serving as the critical point $(T_c, Q_c, \Phi_c)$. It means that the large and small black holes merge into one and coexist. Meanwhile, the isotherm branch undergoes a significant evolution and develops into the critical isotherm $T=T_c$. The isotherm branches passing above the critical point $C$ represent that $Q$ decreases monotonically with increasing $\Phi$. In this case, the facts that $(\frac{\pt Q}{\pt \Phi})_T < 0$ and there is no local minimum or maximum imply that the black hole is thermally stable and no phase transition occurs.

Up to now, it has shown that, in $(Q,\Phi)$ phase plane, critical behaviours of the charged $f(R)$-CIM AdS black hole are analogous to a Van der Waals liquid-vapor system. However, differing from the $P-V$ diagram, both the Van der Waals oscillation and phase transitions occur in the region $T>T_c$. In fact, a higher temperature corresponds to a less charge.

What's more, Fig.\ref{QPhidiag} also displays how the gravity modification parameters $\xi$ and $R_0$ affect the phase structure. Comparing the isotherm $T_{13}$ in Fig.\ref{QP1a1diag} with the isotherm $T_{23}$ in Fig.\ref{QP1b1diag}, as $R_0$ decreases, the unstable stage is shorter. Thus, the large $R_0$ delay the black hole to reach the stable stage. Likewise, for the isotherm $T_{13}$ in Fig.\ref{QP1a1diag} and the isotherm $T_{33}$ in Fig.\ref{QP1a2diag}, we find that large $\xi$ also delay the black hole to reach the stable stage. In addition, taking into account the relation between $\xi$ and $f(R)$ function in Eq.(\ref{eq22}), we can infer that, the smaller value the derivative $f'(R_0)$ takes, the easier the system reaches the stable stage during the phase transition. It is worth pointing out that the charged $f(R)$-CIM AdS black hole (\ref{eq2}) with (\ref{eq22}) would reduce to standard $4d$ RN AdS black hole \cite{Wu:2000id} if specifying $\xi=1$ and replacing $R_0$ with $4\Lambda$ in metric function $N(r)$. Fig.\ref{QP1b2diag} shows us the limit case for $\xi=1$ and demonstrates the consistent critical behaviours in four dimensions with the RN AdS black hole \cite{Ma:2016aat} in Einstein's gravity.

\begin{figure*}[htbp]
  \begin{center}
    \subfigure[~$F$ vs. $T$ for varying $Q$]{\label{FT1diag}\includegraphics[width=2.7in,height=2.3in]{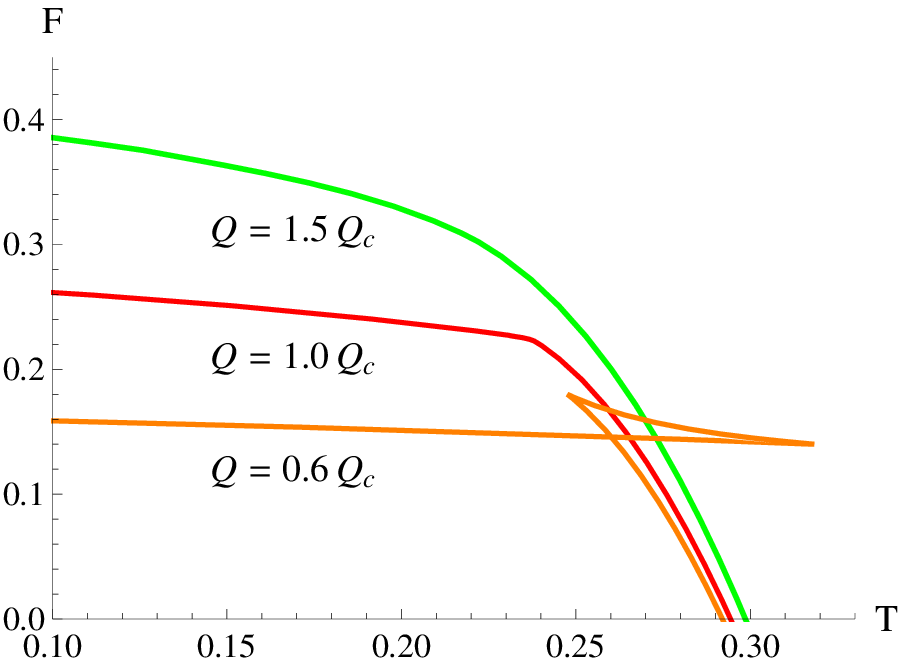}} \quad
    \subfigure[~$F$ vs. $T$ for $Q=0.6Q_c$]{\label{FT2diag}\includegraphics[width=2.7in,height=2.3in]{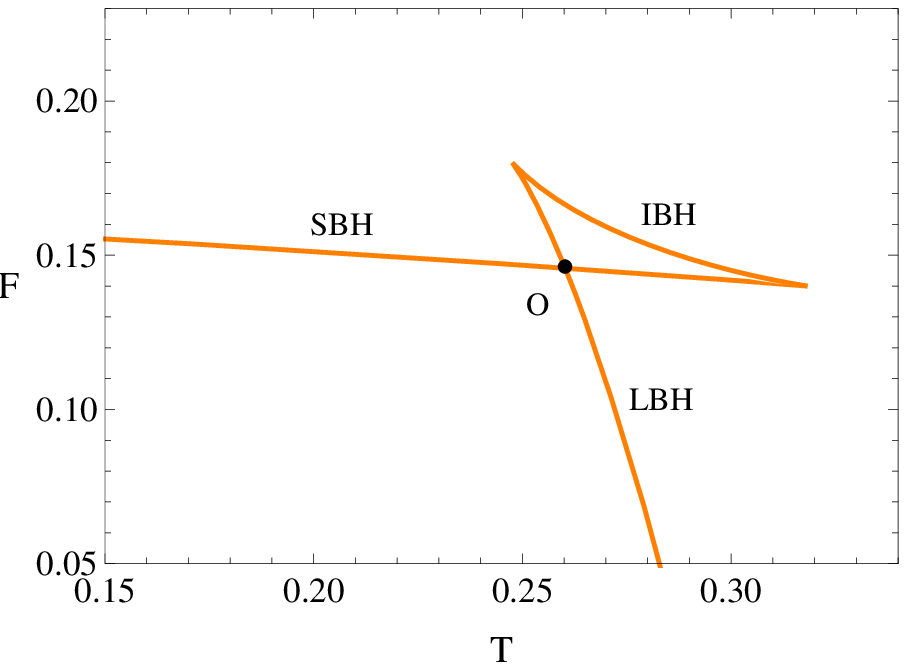}}
    \caption{(color online)\,. Helmholtz free energy $F$ vs. $T$. Here we set $\xi$=$1.5\,,R_0$=$-10$.}
  \label{FTdiag}
  \end{center}
\end{figure*}

The heat capacity $C_Q$ measures the local stability of black holes, while the global thermodynamical properties of physical systems can be reflected by the Helmholtz free energy. In what follows, we will step forward to discuss the Helmholtz free energy defined by
\be
F = M - T S = \frac{\xi}{4} \left(\frac{R_0 r_+^3}{12} + r_+ + \frac{3Q^2}{r_+} \right)\, , \label{eq13}
\ee
where $r_+$ is identified as a function of $Q$ and $T$ according to Eq.(\ref{eq5}). The behaviours of $F$ changing with $T$ for different charge $Q$ are plotted in Fig.\ref{FT1diag}.
Remarkably, similar to a Van der Waals liquid-vapor system, with the charge decreasing below the critical value $Q_c$\,, the free energy $F$ gives rise to a classic ``cusp" catastrophe and displays a characteristic ``swallow tail" which indicates that the black hole phase transition aforementioned is of the first order. Observing the ``swallow tail" more clearly in Fig.\ref{FT2diag}, the system first goes along the small black hole (SBH) branch until the temperature increases to the point $O$. At the point $O$, both the small and the large black holes share the same free energy and can coexist. With the temperature increasing further, the large black hole (LBH) dramatically becomes the preferred thermodynamical state because of lower free energy than the small and the intermediate black holes (IBH). Hence, there is a small-large black hole phase transition at the point $O$. Moreover, due to the relation between $S$ and $r_+$ in Eq.(\ref{eq4}), different horizon radius for the small and the large black holes during the transitions corresponds to the discontinuity of the entropy. That means there is a release of latent heat, and the phase transition at the point $O$ is of the first order.

By now, we have surveyed the local and global properties of the charged $f(R)$-CIM AdS black hole, and investigated its critical behaviours from the macroscopic perspective. The results clearly manifest that the charged $f(R)$-CIM AdS black hole undergoes a first order small-large black hole phase transition, and the critical behaviours qualitatively behave like a Van der Waals liquid-vapor system. By the way, the four dimensional black hole solution (\ref{eq2}) with (\ref{eq22}) degenerates into a standard RN AdS black hole when setting $f'(R_0)=0$ in Eq.(\ref{eq22}). One can find that the critical phenomena above coincide with the case of four dimensional RN AdS black hole \cite{Ma:2016aat} in Einstein-Maxwell theory. Next, we will turn to elaborate the microscopic properties by using the thermodynamic state space geometry.

\section{Thermodynamic state space geometry and microscopic properties} \label{section5}

A systematic statistical description of black hole microstates is still a lack. According to Ruppeiner's proposal, the Riemannian geometry could give insight into the underlying statistical mechanical system. Inspired by the close connection with phase transitions, thermodynamic state space geometry has been extensively adopted for black hole research \cite{Suresh:2014pra,Aman:2005xk,Niu:2011tb,Mirza:2007ev,Lala:2011np,Bellucci:2011gz,Suresh:2014jaa,Mansoori:2013pna,Soroushfar:2016nbu}. To begin with, the metric defined by Ruppeiner \cite{Ruppeiner:1995zz} is given by
\be
g_{i j} = -\frac{\pt^{2}S(x)}{\pt x^{i}\pt x^{j}}\,, \label{eq14}
\ee
in which $x^{i}$ are the extensive variables ascribed to a given thermodynamic system. To evaluate the Ruppeiner metric, one usually proceeds from Weinhold's definition \cite{Weinhold:1975mg} for convenience,
\be
g_{i j}^W = \frac{\pt^{2}U(x)}{\pt x^{i}\pt x^{j}}\,, \label{eq15}
\ee
where $U$ denotes the internal energy of the system. It is demonstrated that the line elements in Ruppeiner geometry and Weinhold geometry are conformally related via the map \cite{Salamon:1983ni,Mrugala:1983ct}
\be
ds^{2}_{R}=\frac{1}{T}ds^{2}_{W}\,, \label{eq16}
\ee
where $T$ is the temperature of black holes. To facilitate calculating, we recast the mass $M$ into a function of entropy $S$ and electric potential $\Phi$ as
\be
M = \frac{\left(12 \pi \Phi^2 - \xi R_0 S + 12 \pi \xi^2\right)\sqrt{S}}{24 \pi^{3/2} \xi^{3/2}}\,. \label{eq17}
\ee
Combining Eqs.(\ref{eq3})-(\ref{eq5}) and (\ref{eq16}), the Ruppeiner metric can be calculated as
\bea
g_{SS}^R &=& \frac{4 \pi  \left(\xi^2-\Phi ^2\right)+\xi R_0 S}{2 S \left(4 \pi  \left(-\xi^2+\Phi ^2\right)+\xi R_0 S\right)}\,, \\ \label{eq18}
g_{S\Phi}^R &=& \frac{8 \pi  \Phi }{4 \pi  \left(-\xi^2+\Phi ^2\right)+\xi R_0 S} = g_{\Phi S}^R\,, \\ \label{eq19}
g_{\Phi \Phi}^R &=& \frac{16 \pi  S}{4 \pi  \left(-\xi^2+\Phi ^2\right)+\xi R_0 S}\,. \label{eq20}
\eea
Programming with Mathematica 10.2, the scalar curvature can be obtained easily. Here we rewrite it as the function of $r_+$ and $Q$,
\be
\mathbb{R}=\frac{8 \Delta }{\pi \xi \left(- R_0 r_+^4 + 4 r_+^2 - 4 Q^2 \right) \left(R_0 r_+^4 + 4 r_+^2 - 12 Q^2\right)^2}, \label{eq21}
\ee
where
\be
\Delta = R_0 r_+^2\left[20 Q^4- \left(3 Q^2-r_+^2\right) \left(6 + R_0 r_+^2\right)r_+^2\right] -8 \left(Q^2-r_+^2\right)^2\,. \nn
\ee
As a result, due to the nonextremal condition (\ref{eq5}), scalar curvature (\ref{eq21}) shares the divergent point with heat capacity (\ref{eq7}). That is to say, the Ruppeiner curvature diverges exactly at the point where the heat capacity diverges, namely, the critical point. In a sense, our research on the charged $f(R)$-CIM AdS black hole proves again the Ruppeiner curvature $\mathbb{R}$ is an excellent measure of the instability of thermodynamic systems \cite{Oshima:1999rs}.

Perhaps more interestingly, the charged $f(R)$-CIM AdS black holes possess the property as ideal Fermi gases.
\begin{figure}[htbp]
\centerline{ \scalebox{0.90}{\includegraphics{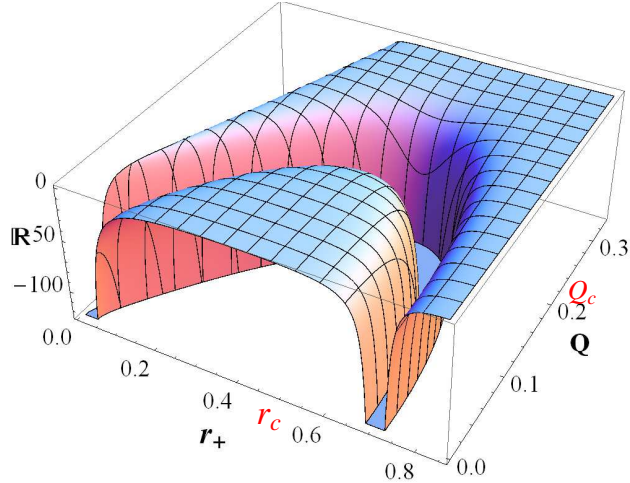}} }
\caption{(color online)\,. Ruppeiner curvature $\mathbb{R}$ varying with $r_+$ and $Q$ for $\xi=1.5$\,, $R_0=-8$. The case for extremal black holes is excluded.}
\label{Rrq3Ddiag}
\end{figure}
As is alluded in Sec.\ref{section1}, Ruppeiner \cite{Ruppeiner:2010tc} declared that different sign of the curvature signals different interaction among the microstructure of thermodynamic systems. For ideal Bose gases, the scalar curvature $\mathbb{R}>0$ and goes to positive infinity at critical point \cite{Janyszek:1990rg,Oshima:1999rs}, and it corresponds to Bose-Einstein condensation. While for ideal Fermi gases, $\mathbb{R}<0$ and diverges to negative infinity at critical point \cite{Janyszek:1990rg,Oshima:1999rs}, which appeared as the Pauli's exclusion principle forbidding two particles to stay in the same state with unlimited repulsive force. As is shown in Fig.\ref{Rrq3Ddiag}, for the charged $f(R)$-CIM AdS black hole, $\mathbb{R}$ always remains nonpositive and negatively diverges, which is amazingly analogous to a Van der Waals liquid-gas system. Just as ideal Fermi gases, $\mathbb{R}<0$ signals that repulsive interaction is permitted in the microscopic system of the charged $f(R)$-CIM AdS black holes. It is noteworthy that, as an important development, our study reveals the microscopic similarity of the charged $f(R)$-CIM AdS black hole with a Van der Waals liquid-gas system. It may serve as a useful guide in deeper understanding the macroscopically similar critical phenomena of the black hole with a Van der Waals liquid-vapor system.

\section{Discussions and conclusions} \label{section6}

Research on black hole phase transitions from more perspectives contributes to understanding black hole thermodynamics more deeply. In this paper, we concentrated on further exploring the phase transitions, critical behaviours and microstructure of the charged $f(R)$-CIM AdS black holes with constant curvature. As a highlight, this study was implemented by employing $(T, Q,\Phi)$ as new state parameters. It provides another perspective other than $(T, P, V)$ description for one to capture more physical pictures of the charged AdS black holes. Interestingly, the charged $f(R)$-CIM AdS black hole exhibits the eye-catching Van der Waals-like phase transition behaviours. What's more, differing from the black holes in Einstein's gravity, we observed that phase structures of the black holes in $f(R)$ theory display an interesting dependence on gravity modification parameters $\xi$ and $R_0$. On the one hand, the critical point is explicitly determined by $\xi$ and $R_0$, on the other hand, both $\xi$ and $R_0$ have an effect on delaying the black holes to reach the stable phase. By the way, the critical behaviours for the limit case $\xi=1$ in four dimensions coincide with those of RN AdS black hole in Einstein's gravity.

Using the thermodynamic state space geometry, we also investigated the microscopic properties of the charged $f(R)$-CIM AdS black holes. The Ruppeiner curvature $\mathbb{R}$ and the heat capacity $C_Q$ diverge exactly at the same point where the small-large black hole phase transition occurs. In a sense, our research on the $f(R)$-CIM AdS black hole proved again that the curvature $\mathbb{R}$ contains the information of black hole phase transition and it indeed serves as an excellent measure of the instability of thermodynamic systems. Perhaps more significantly, we revealed a microscopic similarity between the $f(R)$-CIM AdS black hole and the Van der Waals liquid-vapor system. Remarkably, both their Ruppeiner curvatures $\mathbb{R}$ always remain nonpositive and diverges to negative infinity. This intriguing microscopic similarity may serve as a possible explanation of the macroscopic Van der Waals-like behaviours of the black holes and provide a useful guide in deeper understanding black hole thermodynamics and the process of phase transitions.

\begin{acknowledgments}\vskip -4mm
The authors would like to express sincere gratitude to the anonymous referees for their constructive comments and help in improving this paper greatly. The authors also appreciate the editor for the responsible work. This paper is supported by the National Natural Science Foundation of China under Grant Nos. 11275017 and 11173028.
\end{acknowledgments}

\end{document}